%% file: bruzual.tex
\documentclass[11pt,twoside]{article}

\usepackage{asp2006}
\usepackage{epsf}
\usepackage{graphicx}
\usepackage{lscape}

\begin{document}
\title{Stellar Populations: High Spectral Resolution Libraries. Improved TP-AGB Treatment}
\author{Gustavo Bruzual A.}
\affil{CIDA, AP 264, M\'erida, Venezuela}

\begin{abstract}
I present a short description of the STELIB, HNGSL, IndoUS, MILES,
ELODIE, UVES-POP, and IRTF libraries of empirical stellar spectra
and show some applications of their use in population synthesis models.
When new calculations of the TP-AGB evolutionary phase for stars of different mass
and metallicity are included in population synthesis models, the stellar mass in
galaxies at z from 1 to 3 determined from spectro-photometric data can be up
to 50\% lower than the mass determined from the BC03 models. The ages inferred
for these populations are considerably lower than the BC03 estimates.
\end{abstract}

\section{Introduction}

\input fig1

In a series of conference papers appeared in the past two years, Bruzual
(2004, 2005), I present evolutionary population synthesis models which
are identical in all respects to the \citet{BC03} models, hereafter BC03,
except in the stellar library used. BC03 use the STELIB library compiled
by \citet{STELIB03}.  In these papers I explore models
built with libraries of higher spectral
resolution and which improve upon STELIB on the coverage of the HRD by
including at all metallicities a broader and more complete distribution
of spectral types and luminosity classes.
For reasons of space I present here only a summary of results.
The reader is referred to the previous papers, available electronically,
for details.
The full implementation of the new libraries in the population synthesis
models is in preparation by Charlot \& Bruzual (2007).

STELIB includes observed spectra of 249 stars in a wide range of
metallicities in the wavelength range from 3200 \AA\ to 9500 \AA\ at a
resolution of 3 \AA\ FWHM (corresponding to a median resolving power of
$\lambda / \Delta\lambda \approx 2000$), with a sampling interval of 1 \AA\
and a signal-to-noise ratio of typically 50 per pixel.

HNGSL, the Hubble's New Generation Spectral Library \citep{HL03} contains spectra
for a few hundred stars whose fundamental parameters, including chemical
abundance, are well known from careful analysis of the visual spectrum.
The spectra cover fully the wavelength range from 1700 \AA\ to 10,200 \AA.
The advantage of this library over the ones listed below is the excellent
coverage of the near-UV and the range from 9000 \AA\ to 10,200
\AA, which is generally noisy or absent in the other data sets.

The IndoUS library \citep{VAL04} contains complete spectra over the entire
3460 \AA\ to 9464 \AA\ wavelength region for 885 stars obtained with the
0.9m Coud\'e Feed telescope at KPNO. The spectral resolution is $\approx$
1 \AA\ and the dispersion 0.44 \AA\ pixel$^{-1}$. The library includes
data for an additional 388 stars, but only with partial spectral coverage.
See \citet{GB05} for a discussion of the flux calibration problems in the
IndoUS library.

MILES, the Medium resolution INT Library of Empirical Spectra \citep{MILES06},
contains carefully calibrated and homogeneous quality spectra for 985 stars
in the wavelength range 3525 \AA\ to 7500 \AA\ with 2.3 \AA\ spectral resolution
and 0.9 \AA\ pixel$^{-1}$ sampling. The stars included in this library were
chosen aiming at sampling stellar atmospheric parameters as completely as
possible.

The ELODIE library is a stellar database of 1959 spectra for 1503 stars,
observed with the \'echelle spectrograph ELODIE on the 193 cm telescope at
the Observatoire de Haute Provence. The resolution of this library is
$R = 42,000$ in the wavelength range from 4000 \AA\ to 6800 \AA\
\citep{PAS01A, PAS01B}.
This library has been updated, extended, and used by \citet{LeB04}
in the version 2 of the population synthesis code PEGASE.

The UVES Paranal Observatory Project \citep{UVES03}, has produced a library
of high resolution ($R = \lambda / \Delta\lambda \approx 80,000$) and high
signal-to-noise ratio spectra for over 400 stars distributed throughout
the HRD. For most of the spectra, the typical final SNR obtained in the V band
is between 300 and 500. The UVES POP library is the richest available database
of observed optical spectral lines.

Progress in compiling libraries at IR wavelengths has been slower than in the
optical range. The IRTF Spectral Library, \citet{RVC07}, provides high S/N spectra
for stars covering most of the HR diagram: 333 stars with spectral types W-R
through M, plus 14 L and T stars. The spectra cover the wavelength range
from 0.8 to 4.2 $\mu m$ (in some cases out to 5.0 $\mu m$).
R=2000 for the 0.8-2.5 $\mu m$ range and
R=2500 for the 2.5-5.0 $\mu m$ range. The spectra are corrected for telluric
absorption and then absolutely flux calibrated. In addition, the spectra have
been tied to the 2MASS mags when possible. A good fraction of the data is
available electronically and ready to be used in synthesis models.
Preliminary descriptions of these data can be found in \citet{RJT03} and
\citet{CRV05}.

A parallel effort by
\citet{MQC07} will produce a library of stellar spectra in the K band for
a subset of the stars in the MILES library, insuring an adequate coverage
of stellar physical parameters and non-solar abundance ratios.

There are several on-going projects to improve the existing grids of theoretical
model atmospheres including the computation of high resolution theoretical
spectra for stars whose physical parameters are of interest for population
synthesis. See, for example, \citet{COE07} and references therein.

\section{Use of Different Libraries in Population Synthesis Models}

The 'standard' BC03 reference model represents a simple stellar
population (SSP) computed using the Padova 1994 evolutionary tracks,
the \cite{CHAB03} IMF truncated at 0.1 and 100 M$_\odot$, and either
the STELIB or the BaSeL 3.1, \cite{WES02}, spectral libraries
(see BC03 for details).
For illustration purposes I show in Fig 1 the 13 Gyr spectral energy
distribution (SED) in the optical range for the solar metallicity
standard reference SSP model
computed with the following spectral libraries (top to bottom in decreasing
order of spectral resolution):
\citet{COE07}, IndoUS, MILES, STELIB, HNGSL, \citet{PIC98}, and BaSeL 3.1.

The higher resolution models do not provide new information in what respects
to color or color evolution, i.e., all the SEDs in Fig 1 have the same
overall shape.
The major advantage of using high spectral resolution models is to study
absorption features. The behavior of line strength indices
defined in lower resolution spectra can be explored in detail in the
higher resolution SEDs. In some instances, e.g some of the Lick indices,
the high resolution spectra show clear evidence that the wavelength intervals
defining these indices should be revised because of contamination by
other chemical elements. More important, the high resolution spectra provide
the opportunity to define new line strength indices that measure the
intensity of absorption lines that are unnoticeable in low resolution
spectra.

\input fig2
\input fig3

Fig 2 shows the behavior in time of several new spectral indices
defined by \citet{MJS07} upon inspection of the IndoUS version of the
BC03 models. For ages above 5 Gyr any of these indices is a good indicator
of the metallicity of the stellar population. When combined with
indices that are good age indicators, such as the I$_{200}$ and I$_{275}$
indices defined by \citet{VA99}, we obtain close to orthogonal index-index diagrams
in which the age-metallicity degeneracy is broken and can be used to
establish approximate ages for the stellar populations in early-type
galaxies, cf. Fig. 3. The data points in Fig. 3 represent the values of
these indices measured in very high S/N coadded SDSS spectra of early-type
galaxies (J. Brinchmann, private communication)
with velocity dispersion close to the value indicated in each frame.
A tendency is observed for most massive galaxies being older and more metal
rich.

\subsection{Improved TP-AGB treatment}

\input fig4

It has been pointed out by several authors, e.g. \citet{CM06}, \citet{KG07},
that the estimates of the age and mass of the stellar population present in a galaxy
depend critically on the ingredients of the stellar population model
used to fit the galaxy spectrum. \citet{CM06} have shown that the treatment 
of the thermally pulsing asymptotic giant branch (TP-AGB) phase of stellar
evolution is a source of major discrepancy in the determination
of the spectroscopic age and mass of high-z $(1.4<z<2.7)$ galaxies.
The mid-UV spectra of these galaxies indicate ages in the range from
0.2-2 Gyr, at which the contribution of TP-AGB stars in the rest-frame
near-IR sampled by Spitzer is expected to be at maximum. \citet{CM06} find that
in general the \citet{CM05} models (M05 hereafter) provide better fits than the BC03
and other models available in the literature, and indicate systematically lower ages
and, on average, 60\% lower masses for the stellar populations sampled in these galaxies.
According to \citet{CM06} the source of this discrepancy is primarily a consequence
of the different treatment of the TP-AGB phase in the evolutionary models.

\citet{PM07} have recently concluded new calculations of the TP-AGB evolutionary
phase for stars of different mass and metallicity. The evolution of the stars is
now computed accounting for the changes in the chemical composition of the envelopes.
As a consequence of this prescription, the signature of TP-AGB stars around 1 Gyr,
i.e the red color of the integrated stellar population, becomes more important.
Fig. 4 compares the B-V and V-K color evolution of models computed using the
\citet{PM07} TP-AGB evolutionary tracks (Bruzual \& Charlot 2007, CB07 hereafter)
with the same for the BC03 and M05 models. In B-V the CB07 and BC03 models are
identical at all ages. At early and late ages both sets of models have the
same V-K color. At intermediate ages the CB07 models are considerably redder than
the BC03 models. At late ages, the BC03 and CB07 models match very well the
observations of nearby early-type galaxies, whereas the M05 models are too blue.

\input fig5

Fig. 5 compares the evolution of the fraction of the K-band luminosity
emitted by TP-AGB stars in a solar metallicity SSP for the CB07 and the BC03
models. At maximum, the TP-AGB contributes 60\% of the K-light in the CB07 model
but only 40\% in the BC03 model. The peak emission in the BC03 model occurs
at around 1 Gyr whereas in the CB07 model it stays high and close to constant
from 0.1 to 1 Gyr. The bottom frame of Fig. 5 shows that the stellar mass determined
from the CB07 model can be up to 50\% lower than the mass determined from the BC03
model. The lower CB07 values are in agreement with the masses determined by \citet{CM06}.
See \citet{GB06} and \citet{CB07} for details.

\acknowledgements 

I thank Cesare Chiosi for his contribution to making all of this possible,
Paola Marigo and Leo Girardi for providing their calculations of the TP-AGB 
evolutionary phase ahead of publication, and St\'ephane Charlot for allowing me
to show results of a joint paper in preparation.

\input bibliography
\end{document}

%% file: fig1.tex
\begin{figure}[!ht]
\begin{center}
\includegraphics[scale=0.61]{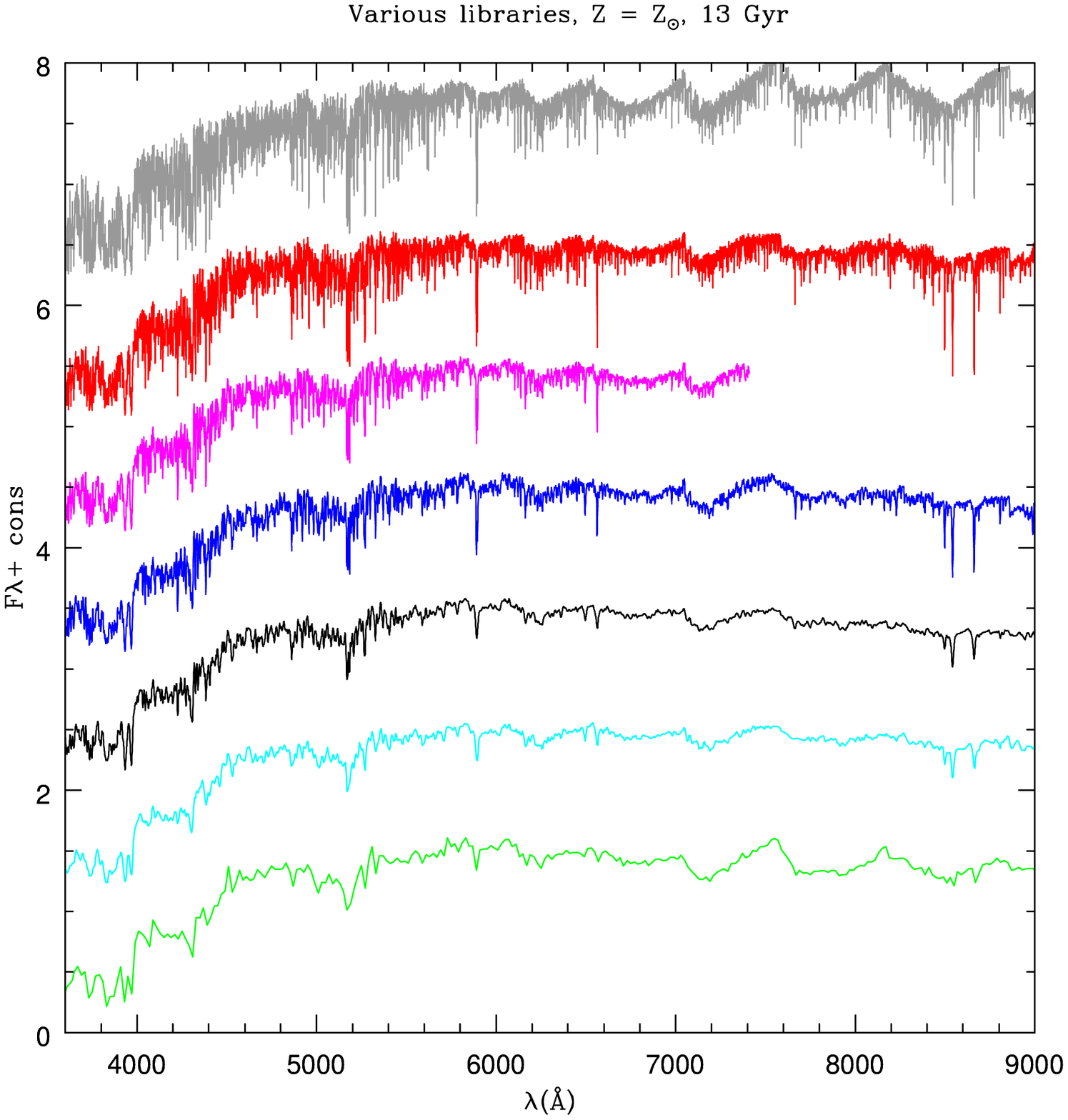}
\end{center}
\caption{
  BC03 standard SSP model spectra for solar metallicity at 13 Gyr in
  the wavelength range from 3600 \AA\ to 9000 \AA\ computed with the
  following spectral libraries (top to bottom in decreasing order of
  spectral resolution): Coelho et al. (spectral resolution $< 1$ \AA),
  IndoUS (1 \AA), MILES (2.3 \AA), STELIB (3 \AA), HNGSL (5 \AA),
  Pickles (5 \AA), and BaSeL 3.1 (20 \AA).
  The spectra have been scaled and shifted in the vertical direction for clarity.
}
\end{figure}

%% file: fig2.tex
\begin{figure}[!ht]
\begin{center}
\includegraphics[scale=0.54]{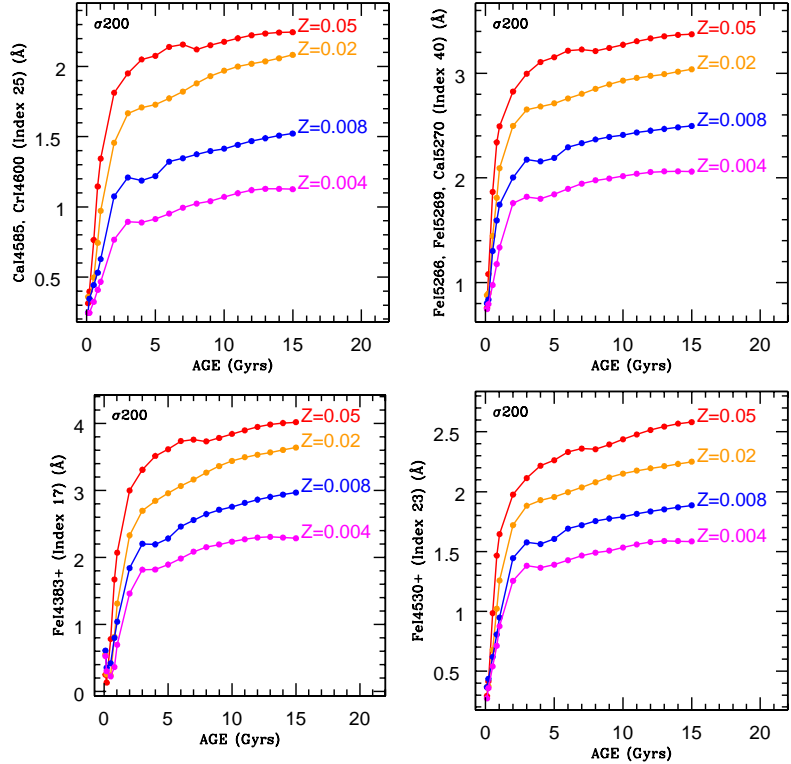}
\end{center}
\caption{
Behavior in time of several new spectral indices defined by \citet{MJS07}
upon inspection of the IndoUS version of the BC03 models. For ages above
5 Gyr any of these indices is a good indicator of the metallicity of the
stellar population. The indices were measured in the model spectra at a
resolution corresponding to a velocity dispersion $\sigma = 200$ km s$^{-1}$.
}
\end{figure}

%% file: fig3.tex
\begin{figure}[!ht]
\begin{center}
\includegraphics[scale=0.6]{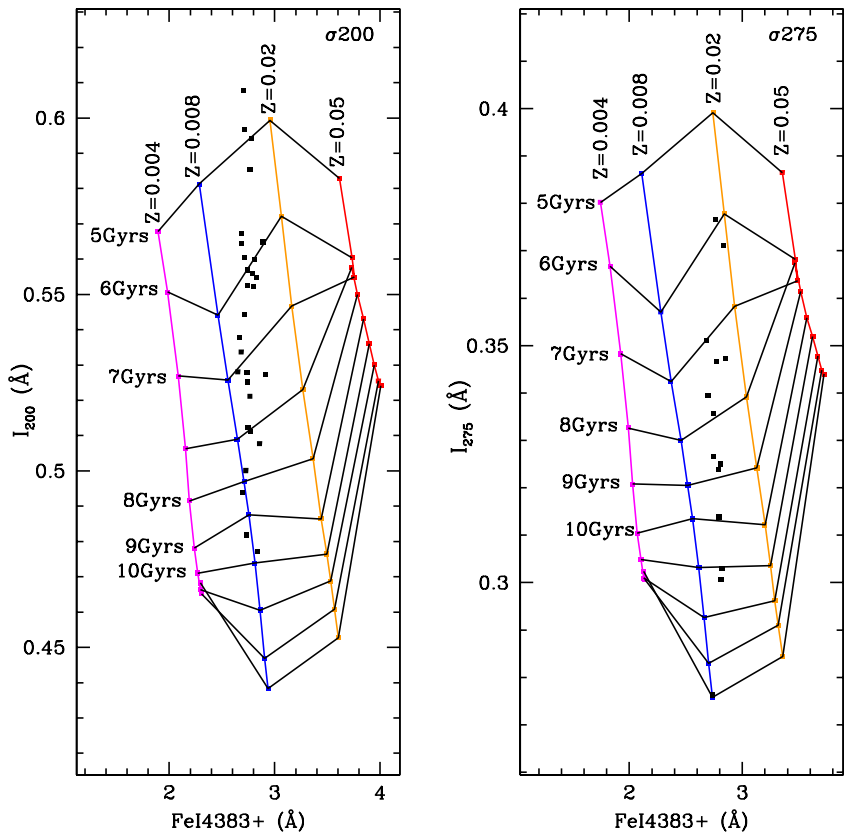}
\end{center}
\caption{
Close to orthogonal index-index diagrams obtained when combining
the robust age indicator indices I$_{200}$ and I$_{275}$ defined
by \citet{VA99} with the metallicity indicator index measuring FeI 4383
(index No 17 shown in Fig. 2) defined by \citet{MJS07}.
The indices were measured at a resolution corresponding
to a velocity dispersion $\sigma = 200$ km s$^{-1}$ (LHS frame) and
275 km s$^{-1}$ (RHS frame). The data points represent the indices
measured in very high S/N coadded SDSS spectra of early-type galaxies
(J. Brinchmann, private communication)
with velocity dispersion close to the value indicated in each frame.
}
\end{figure}

%% file: fig4.tex
\begin{figure}[!ht]
\begin{center}
\includegraphics[scale=0.5]{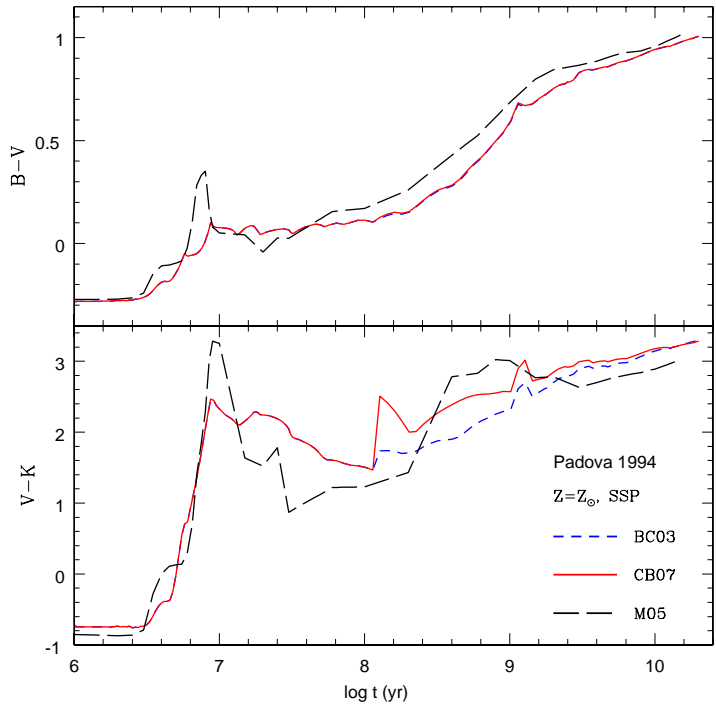}
\end{center}
\caption{
Comparison of the B-V and V-K color evolution of the CB07 models (solid line)
computed using the \citet{PM07} TP-AGB evolutionary tracks with the BC03
(short-dashed line) and the M05 (long-dashed line) models.
}
\end{figure}

%% file: fig5.tex
\begin{figure}[!ht]
\begin{center}
\includegraphics[scale=0.54]{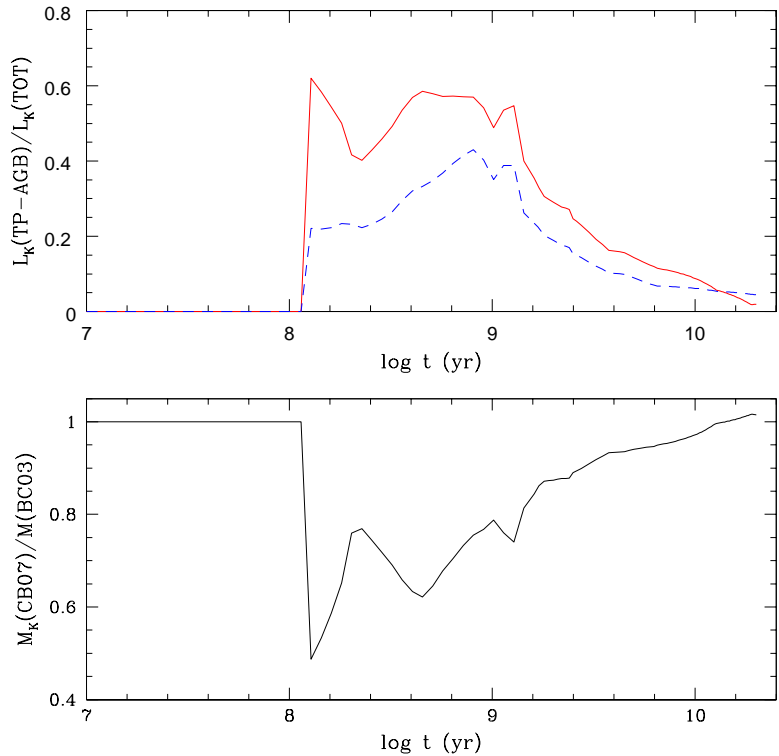}
\end{center}
\caption{
 Top frame: time dependence of the fraction of the K-band luminosity
 emitted by TP-AGB stars in a solar metallicity SSP for the CB07
 (solid line) and the BC03 (dashed line) models.
 Bottom frame: ratio of the stellar mass determined from the CB07 and the
 BC03 models for a given K-band galaxy luminosity.
}
\end{figure}